\documentclass[sigconf]{acmart}
\usepackage{graphicx}
\usepackage{booktabs}
\usepackage{adjustbox}

\AtBeginDocument{%
  }

\setcopyright{none}
\copyrightyear{}
\acmYear{2025}
\acmDOI{}
\acmConference[CS7DS1 - Data Analytics]{}{May 2025}{Trinity College Dublin}
\acmISBN{}

\begin{document}

\title[Copula Analysis of Risk]{Copula Analysis of Risk: A Multivariate Risk Analysis for VaR and CoVaR using Copulas and DCC-GARCH}

\author{Aryan Singh}
\email{singha12@tcd.ie}
\affiliation{%
    \country{}
}

\author{Paul O'Reilly}
\email{oreilp11@tcd.ie}
\affiliation{%
    \country{}
}

\author{Daim Sharif}
\email{sharifd@tcd.ie}
\affiliation{%
    \country{}
}

\author{Patrick Haughey}
\email{haughepa@tcd.ie}
\affiliation{%
    \country{}
}

\author{Eoghan McCarthy}
\email{mccare23@tcd.ie}
\affiliation{%
    \country{}
}

\author{Sathvika Thorali Suresh}
\email{thoralis@tcd.ie}
\affiliation{%
    \country{}
}

\author{Aakhil Anvar}
\email{anvara@tcd.ie}
\affiliation{%
    \country{}
}

\author{Adarsh Sajeev Kumar}
\email{sajeevka@tcd.ie}
\affiliation{%
    \country{}
}

\renewcommand{\shortauthors}{Aryan, Paul, Daim, Patrick, Eoghan, Sathvika, Akhil, Adarsh}

\begin{abstract}
A multivariate risk analysis for VaR and CVaR using different copula families is performed on historical financial time series fitted with DCC-GARCH models. A theoretical background is provided alongside a comparison of goodness-of-fit across different copula families to estimate the validity and effectiveness of approaches discussed.
\end{abstract}

\maketitle

\section{Introduction}

The accurate quantification of risk in modern financial markets has never been more critical. Traditional risk measures such as Value-at-Risk (VaR) capture only marginal tail risk, overlooking the complex interdependencies that can amplify shocks across assets or institutions. Conditional Value-at-Risk (CoVaR) addresses this shortcoming by measuring the risk of one portfolio conditional on another being in distress, thereby providing a natural metric for systemic risk assessment. In particular, CoVaR has emerged as a cornerstone in regulatory stress testing and portfolio risk management, as it explicitly accounts for contagion effects during periods of market turbulence \cite{adrian2016covar, basel2018covar}.

To model both volatility clustering at the individual asset level and time-varying dependence across assets, the copula-DCC-GARCH framework offers a powerful and flexible paradigm. This framework has been studied and effectively utilised in its application to financial time series \cite{engle2002dcc, patton2006copula, hafner2012copula, harvey2013dynamic, creal2013generalized}. In this approach, each return series is first filtered through a univariate GARCH process to capture conditional heteroskedasticity, yielding standardized residuals that are then linked via a Dynamic Conditional Correlation (DCC) model. The resulting time-varying correlation matrices serve as inputs to copula functions to assemble a full joint distribution while preserving arbitrary marginal behaviors. By decoupling marginal dynamics from dependence structure, this framework accommodates a rich variety of marginal distributions and copula families, including heavy-tailed Student-t and asymmetric Archimedean types.

Despite its theoretical appeal, empirical applications often rely on ad-hoc choices of copula families or calibration methods, and a systematic comparison of their relative merits in CoVaR estimation remains underexplored. In particular, questions persist regarding (i) the impact of tail dependence properties of different copulas on CoVaR forecasts, and (ii) the trade-offs between goodness-of-fit and predictive reliability in a dynamic setting. To address these gaps, we construct a controlled synthetic data experiment in which we benchmark Gaussian, Student-t, and Clayton copulas within the DCC-GARCH pipeline. Goodness-of-fit is evaluated via information-criteria and distance-based tests.

Our contributions are threefold. First, we provide a reproducible pipeline for synthetic time-series generation, marginal GARCH calibration, DCC estimation, and copula coupling. Second, we deliver a head-to-head evaluation of copula families, quantifying how tail-dependence characteristics translate into CoVaR accuracy. Third, we offer practical guidance on model selection for risk practitioners, balancing statistical fit against the demands of real-time risk monitoring. The remainder of this report is structured as follows. Section 2 reviews the theoretical underpinnings of distributions, copulas, and time-series models. Section 3 details our data-generation and evaluation methodologies. Section 4 presents the results of our comparative study. Finally, Section 5 concludes with a summary of work done and future improvements to our approach.

\section{Background}

\subsection{Distributions}

The following definitions have been adapted from standard probability textbooks \cite{casella2002, degroot2012}.

\subsubsection{Cumulative Distribution Functions}

A random variable \( X \) defined on a subset \( N \) of the real numbers is \textit{continuous} if there exists a cumulative distribution function (CDF) \( F_X: N \subseteq \mathbb{R} \to [0, 1] \) such that:
\[
F_X(x) = \mathbb{P}(X \le x) \quad \forall x \in N
\]
If the random variable \( X \) is \textit{absolutely continuous} then there exists a probability density function (PDF) \( f_X: N \subseteq \mathbb{R} \to \mathbb{R}^+ \) such that:
\[
F_X(x) = \int_{-\infty}^t f_X(t)\space dt
\]
Equivalently assuming continuity of \( f_X \) we have:
\[
f_X(x) = \frac{dF_X(x)}{dx}
\]

\subsubsection{Probability Integral Transform}

Let \( X \) be a continuous random variable with CDF \( F_X \). Then the \textit{probability integral transform} is:
\[
U = F_X(X) \sim \mathrm{U}(0, 1)
\]
This transformation preserves probability:
\[
\mathbb{P}(U \leq u) = u \quad \forall u \in [0, 1]
\]
This transform standardizes any continuous variable to a Uniform(0,1) scale, enabling a common foundation for copula-based dependence modeling.

\subsubsection{Joint Distributions}

Let \( X_1,\ldots,X_n \) be continuous random variables that are all defined on the same probability space (e.g. subsets of the real numbers \( N_1,\ldots,N_n \)). 

Then, the variables \( X_1,\ldots,X_n \) can be considered \textit{jointly distributed} with a joint CDF \( F_{X_1,\ldots,X_n}: N_i \subseteq \mathbb{R}^n \to [0, 1]^n \) such that:
\[
F_{X_1,\ldots,X_n} = \mathbb{P}(X_1 \le x_1,\ldots,X_n \le x_n) \quad \forall x_i \in N_i, \forall i \in 1,\ldots,n
\]

The PDF \( f_{X_1,\ldots,X_n} \) of a continuous random vector \((X_1, \ldots, X_d)\) is defined as the mixed partial derivative of its CDF \( F_{X_1,\ldots,X_n} \). For \( (x_1, \ldots, x_n) \in \mathbb{R}^d\), it is given by:
\[
f_{X_1,\ldots,X_n}(x_1, \ldots, x_n) = \frac{\partial^n F_{X_1,\ldots,X_n}(x_1, \ldots, x_n)}{\partial x_1 \cdots \partial x_n},
\]
The PDF exists if \( F \) is absolutely continuous.

\subsubsection{Marginal Distributions}

Let \( X_1,\ldots,X_n \) be continuous random variables with joint CDF \( F_{X_1,\ldots,X_n}(x_1,\ldots,x_n) \), with each random variable \( {X_i} \) taking values on \( N_i = [a_i, b_i] \quad \forall i \in {1,..,n}\). 

Then, the \textit{marginal distribution} for the random variable \( {X_i} \) has a CDF given by:
\[
F_{X_i}(x_i) = F_{X_1,\ldots,X_n}(b_1,\ldots,b_{i-1},x_i,b_{i+1},\ldots,b_n)
\]
Assuming the marginal PDF exists, it is given by:
\[
f_{X_i}(x) = \frac{dF_{X_i}(x)}{dx}
\]

In the bi-variate case the CDF formulation simplifies to the variables \( X_1,X_2 \) taking values on \( [a_1, b_1] \times [a_2,b_2] \) with joint CDF \( F_{X_1,X_2}(x_1,x_2) \) having marginals:
\[F_{X_1}(x_1) = F_{X_1,X_2}(x_1,b_2) \quad \text{and} \quad F_{X_2}(x_2) = F_{X_1,X_2}(b_1,x_2)\]

\subsubsection{Elliptical Distributions}

The \textit{characteristic function} \(\varphi : \mathbb{R}^n \to \{z\in\mathbb{C}^n : |z| < 1\}\) of a random variable \( X \) (or vector of random variables) is defined as follows:
\[
\varphi_X(t) = \mathbb{E}[e^{itX}]
\]
If the PDF exists, the characteristic function can also be obtained by applying the Fourier transform to the PDF, i.e.
\[
\varphi_X(t) = \int_{-\infty}^{\infty} f_X(x)e^{-2\pi ixt} dx
\]
The random variable \( X_1,\ldots,X_n \) have an \textit{elliptical distribution} if the characteristic function \(\varphi\) can be represented as follows:
\[
\varphi_{X}(x) = e^{ix\cdot\mu}\psi(x^T\Sigma x)
\]
where $\mu \in \mathbb{R}^n$ is a centrality vector 
(representing mean/median depending on the context), $\psi : \mathbb{R}^+ \to \mathbb{R} $ is a normalised positive definite function and $\Sigma$ is a scale matrix (representing correlation/covariance depending on the context).

Some common examples of elliptical distributions are the multivariate gaussian and student-t distributions.

\subsection{Copulas}

\subsubsection{Definition \cite{sklar1959}}

Let \( X_1,\ldots,X_n \) be continuous random variables with joint CDF \( F_{X_1,\ldots,X_n}(x_1,\ldots,x_n) \). By the probability integral transform, the random vector \( (U_1,\ldots,U_n) 
 = (F_{X_1}(X_1),\ldots,F_{X_n}(X_n))  \) comprised of the marginal CDFs has a standard uniform distribution on \( [0,1]^n \)

 The copula \( C \) of the vector \( (X_1,\ldots,X_n) \) is defined as the joint CDF of the vector \( (U_1,\ldots,U_n) \), namely:
 \[
 C(u_1,\ldots,u_n) = \mathbb{P}(U_1 \le u_1,\ldots,U_n \le u_n) \quad \forall u_i \in [0,1], \space \forall i \in 1,\ldots,n
 \]
 The copula density $c$, if it exists, is the joint PDF of the vector \( (U_1,\ldots,U_n) \), which satisfies:
\[
c(u_1, \ldots, u_n) = \frac{\partial^n C(u_1, \ldots, u_n)}{\partial u_1 \cdots \partial u_n}.
\]
Due to continuity of \(F_{X_i}\), the inverse CDFs \(F^{-1}_{X_i}\) can be used to provide an alternative definition for the copula \( C \): 
 \[
 C(u_1,\ldots,u_n) = \mathbb{P}(X_1 \le F^{-1}_{X_1}(u_1),\ldots,X_n \le F^{-1}_{X_n}(u_n))
 \]
 \[
 \forall u_i \in [0,1], \space \forall i \in 1,\ldots,n
 \]

In essence, a copula is the function that ''glues'' individual marginals into a joint distribution, isolating pure dependence from the shape of each marginal.

\subsubsection{Sklar's Theorem \cite{sklar1959}}

Let \( X_1,\ldots,X_n \) be continuous random variables with joint CDF \( H(x_1,\ldots,x_n) \) and marginal CDFs \( F_i(x_i) \). Then, there exists a copula \( C \) such that:
\[
H(x_1,\ldots,x_n) = C(F_1(x_1),\ldots,F_n(x_n))
\]

Conversely, given a copula \( C \) and marginal CDFs \( F_i(x_i) \), then \( C(F_1(x_1),\ldots,F_n(x_n))\) yields a joint CDF of dimension \( n \) with marginals \( F_i \).

This theorem also implies the decomposition of the joint PDF into marginal and copula density components, i.e:
\[
h(x_1,\ldots,x_n) = c(F_1(x_1),\ldots,F_n(x_n))\cdot\prod_{i=1}^n f_{i}(x_i)
\]
where $f_i$ are the marginal PDFs, $F_i$ are the marginal CDFs, $c$ is the copula density and $h$ is the joint PDF.

\subsubsection{Elliptical Copulas \cite{joe2014}}

Elliptical copulas are those that are generated from the CDF of an elliptical distribution, as follows:
\[
C_{\Sigma}(u) = F_{\Sigma}(F_1^{-1}(u_1),\ldots,F_n^{-1}(u_n))
\]
where $F_{\Sigma}$ is the CDF of an elliptical distribution with scale matrix $\Sigma$, and $F_i^{-1}$ are quantile functions of the marginals.

Elliptical copulas inherit symmetry and tail-behavior from their parent distributions (e.g. Gaussian, Student-t), making them well-suited for modeling roughly balanced co-movements.

\subsubsection{Archimedean Copulas \cite{nelsen2006}}

Archimedean copulas are a family of copulas generated by a completely monotonic generator function $\varphi: [0,1] \to [0,\infty]$ generally parameterised by a single real parameter \(\theta\) in the following fashion:
\[
C(u;\theta) = \varphi^{-1}\left(\varphi(u_1;\theta) + \cdots + \varphi(u_n;\theta)\right)
\]
where $\varphi^{-1}$ is the pseudo-inverse satisfying
\[
\varphi^{-1}(\varphi(t;\theta)) = t \quad \forall t \space \forall \theta
\]

Built from a generator function, Archimedean copulas flexibly capture asymmetric dependence, especially useful when extremes in one tail dominate.

\subsubsection{Empirical Copulas \cite{genest1995}}

For $k$ observations in $n$ dimensions $(\{{X_1}^{(i)}\}_{i=1}^n,\ldots,{X_n}^{(i)}\}_{i=1}^k)$, the \textit{empirical copula} is defined as:
\[
C_k(u) = \frac{1}{k}\sum_{i=1}^k \prod_{j=1}^n \mathbb{I}\left(\frac{\mathrm{R}_j^{(i)}}{n+1} \leq u_j\right)
\]
where $\mathrm{R}_j^{(i)}$ is the rank of $X_j^{(i)}$ among $\{X_j^{(1)},\ldots,X_j^{(k)}\}$ and \(\mathbb{I}\) is the indicator function.

Empirical copulas derive dependence directly from data ranks, offering a nonparametric snapshot free of any predetermined functional form.

\subsection{Copula Families}

\subsubsection{Gaussian Copula \cite{joe2014}}

\begin{figure}[h]
    \centering
    \includegraphics[width=0.4\textwidth]{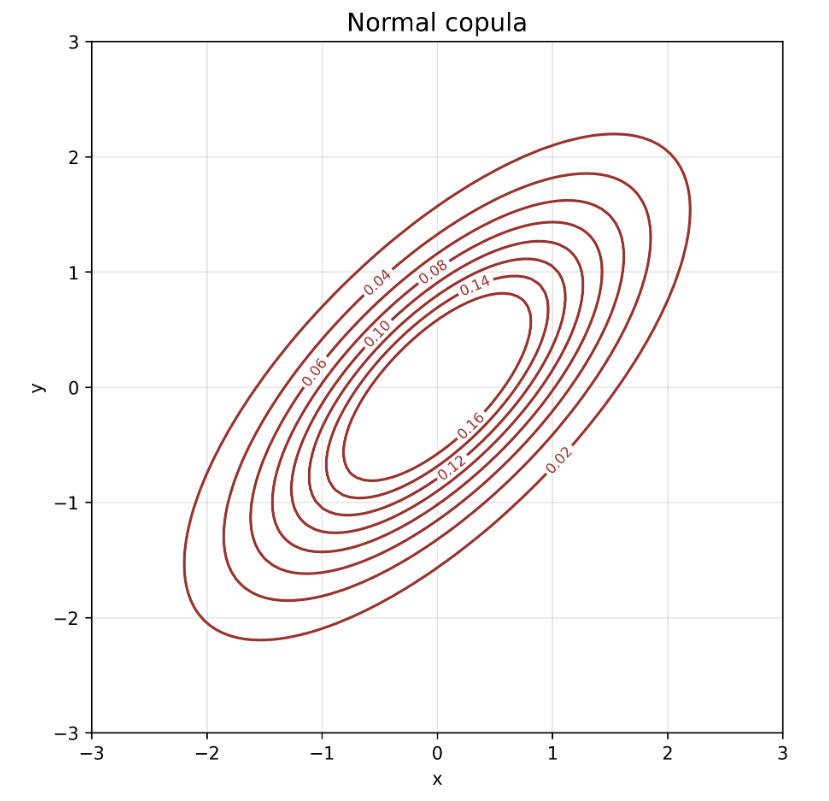}
    \caption{Contour plot of the Gaussian copula}
    \label{fig:gumbel}
\end{figure}

The \textit{Gaussian copula} is an elliptical copula derived from the multivariate normal distribution. For a correlation matrix \(\mathbf{\Sigma} \in [-1,1]^{n \times n}\) and marginals \(u_1, \ldots, u_n \in [0,1]\), it is defined as:
\[
C_{\text{Gauss}}(u_1, \ldots, u_n) = \Phi_{\mathbf{\Sigma}}( \Phi^{-1}(u_1), \ldots, \Phi^{-1}(u_n)),
\]
where \(\Phi_{\mathbf{\Sigma}}\) is the joint CDF of a multivariate normal distribution with mean \(\mathbf{0}\) and correlation matrix \(\mathbf{\Sigma}\), and \(\Phi^{-1}\) is the inverse CDF of the univariate standard normal distribution.

\subsubsection{Student-t Copula \cite{joe2014}}

\begin{figure}[h]
    \centering
    \includegraphics[width=0.4\textwidth]{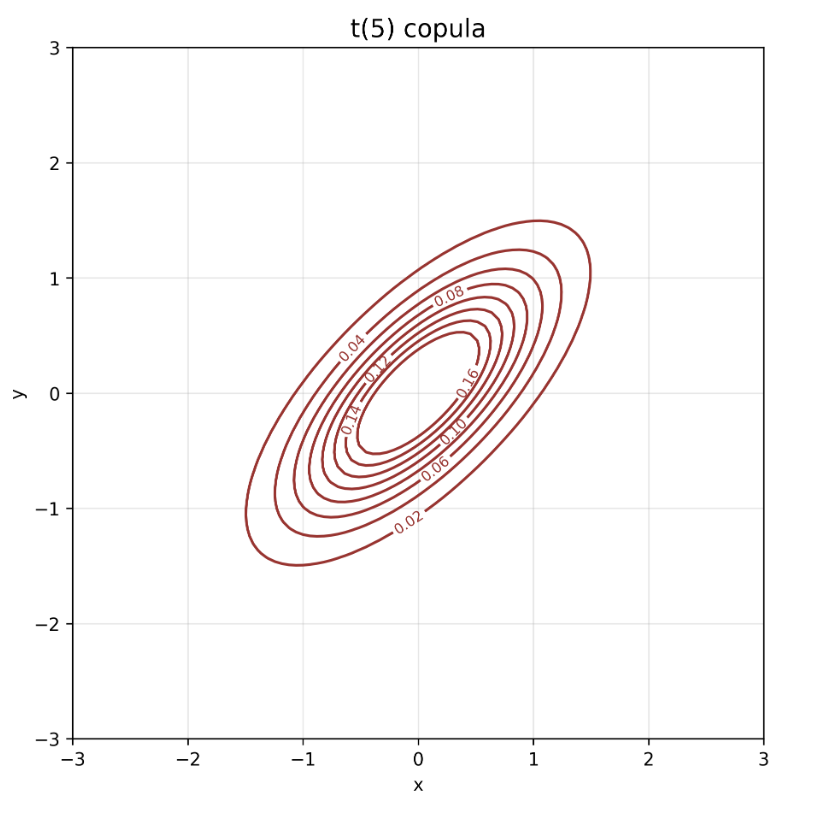}
    \caption{Contour plot of the Student-t copula with $\nu = 5$}
    \label{fig:gumbel}
\end{figure}

The \textit{Student-t copula} is an elliptical copula derived from the multivariate Student-t distribution. For degrees of freedom \(\nu > 0\), correlation matrix \(\mathbf{\Sigma}\), and marginals \(u_1, \ldots, u_n \), it is defined as:
\[
C_{t}(u_1, \ldots, u_n;\nu) = t_{\nu, \mathbf{\Sigma}}\left( t_{\nu}^{-1}(u_1), \ldots, t_{\nu}^{-1}(u_n) \right),
\]
where \(t_{\nu, \mathbf{\Sigma}}\) is the joint CDF of a multivariate Student-t distribution with \(\nu\) degrees of freedom and correlation matrix \(\mathbf{\Sigma}\), and \(t_{\nu}^{-1}\) is the inverse CDF of the univariate Student-t distribution with \(\nu\) degrees of freedom.

\subsubsection{Clayton Copula \cite{nelsen2006}}

The \textit{Clayton copula} is an Archimedean copula with generator function \(\varphi(t) = \frac{t^{-\theta} - 1}{\theta}\) for \(\theta > 0\). For \(u_1, \ldots, u_n \in [0,1]\), it is expressed as:
\[
C_{\text{Clayton}}(u_1, \ldots, u_n;\theta) = \left( \sum_{i=1}^n u_i^{-\theta} - n + 1 \right)^{-1/\theta}
\]
The parameter \(\theta\) governs lower tail dependence, with higher \(\theta\) corresponding to stronger lower tail dependence.

\begin{figure}[h]
    \centering
    \includegraphics[width=0.4\textwidth]{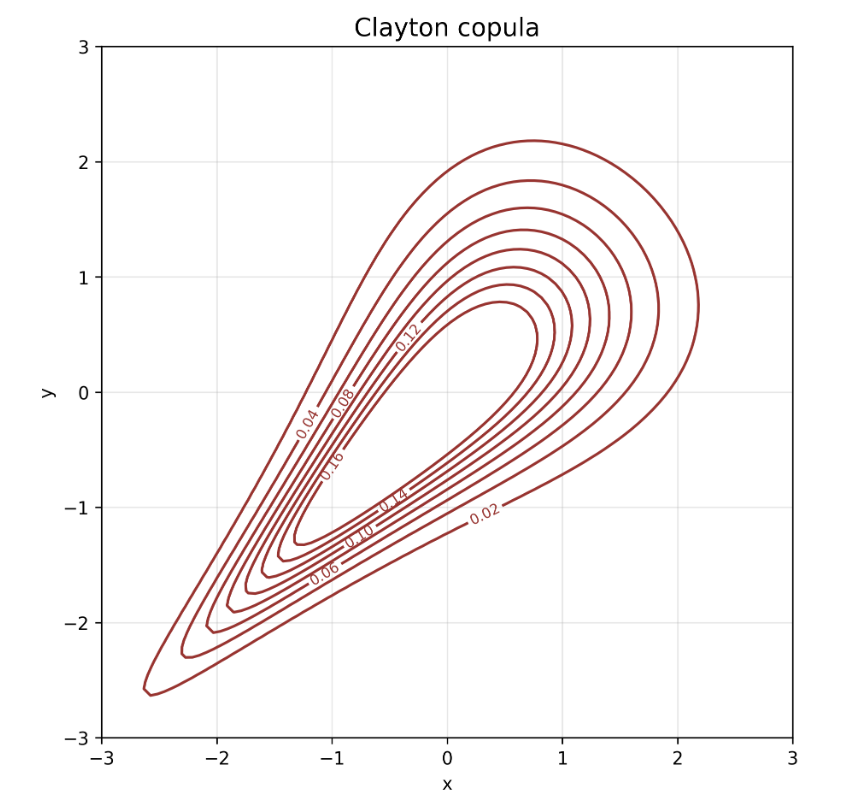}
    \caption{Contour plot of the Clayton copula with $\theta=2$}
    \label{fig:gumbel}
\end{figure}

\subsubsection{Gumbel Copula \cite{nelsen2006}}

\begin{figure}[h]
    \centering
    \includegraphics[width=0.4\textwidth]{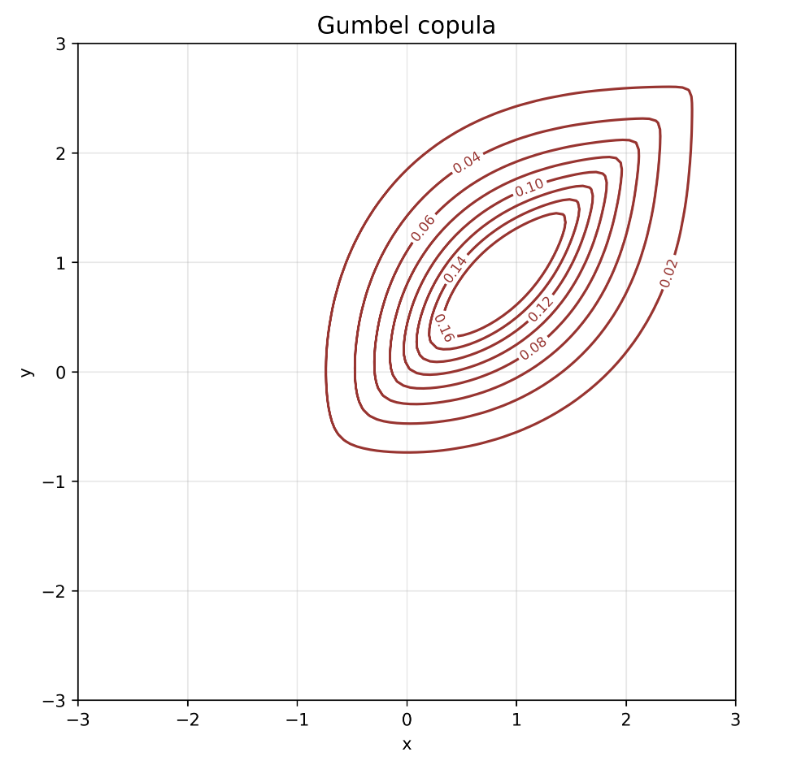}
    \caption{Contour plot of the Gumbel copula with $\theta=5$}
    \label{fig:gumbel}
\end{figure}

The \textit{Gumbel copula} is an Archimedean copula with generator function \(\varphi(t) = (-\ln t)^\theta\) for \(\theta \geq 1\). For \(u_1, \ldots, u_n \in [0,1]\), it is defined as:
\[
C_{\text{Gumbel}}(u_1, \ldots, u_n;\theta) = \exp\left( -\left( \sum_{i=1}^n (-\ln u_i)^\theta \right)^{1/\theta} \right)
\]
The parameter \(\theta\) controls upper tail dependence, with higher \(\theta\) corresponding to stronger upper tail dependence.

\subsection{Time Series}

\subsubsection{Definition \cite{hamilton1994}}

A \textit{time series} is a sequence of random variables $\{X_t\}$ indexed by time $t \in \mathbb{Z}$, typically observed at discrete intervals:
\[
\{X_t\} = \{\ldots, X_{-1}, X_0, X_1, \ldots, X_T\}\quad \text{(for sample size $T$)}
\]
where the temporal ordering contains potential dependence structures.

Time series respect the order of observations, capturing how past values influence future outcomes—an essential feature when modeling financial returns.

\subsubsection{Stationarity \cite{hamilton1994}}

A time series \( \{X_t\} \) is \textit{strictly stationary} if:
\[
F_{X,\ldots,X_{t_k}}(x_1,\ldots,x_k) = F_{X_{t_1+h},\ldots,X_{t_k+h}}(x_1,\ldots,x_k)\quad \forall h,k,t
\]
A time series \( \{X_t\} \) is \textit{weakly stationary} if:
\begin{align*}
    \mathbb{E}[X_t] &= \mu \in \mathbb{R} \quad (\text{constant}) \\
    \mathrm{Var}(X_t) &= \sigma^2 \in \mathbb{R} \quad (\text{constant}) \\
    \mathrm{Cov}(X_t, X_{t+h}) &= \gamma(h) \quad (\text{depends only on lag $h$})
\end{align*}

Stationarity ensures that mean, variance, and auto-covariances remain stable over time, a prerequisite for reliable estimation and inference.

\subsubsection{ARCH(p) (Autoregressive Conditional Heteroskedasticity) \cite{engle1982}}

For innovations $\varepsilon_t = \sigma_t z_t$ where $z_t \sim \mathrm{IID}(0,1)$:
\[
\sigma_t^2 = \omega + \sum_{i=1}^p \alpha_i \varepsilon_{t-i}^2,\quad \omega > 0,\ \alpha_i \geq 0
\]
Key property: Volatility clustering through persistence parameter $\sum \alpha_i < 1$.

ARCH models explain volatility clustering, i.e. periods of high variance following high variance, by letting current variance depend on past squared shocks.

\subsubsection{GARCH(p,q) (Generalised ARCH) \cite{bollerslev1986}}

Extends ARCH with autoregressive volatility terms:
\[
\sigma_t^2 = \omega + \sum_{i=1}^p \alpha_i \varepsilon_{t-i}^2 + \sum_{j=1}^q \beta_j \sigma_{t-j}^2 \quad \omega > 0,\ \alpha_i,\beta_j \geq 0
\]
Stationarity requires $\sum_{i=1}^p \alpha_i + \sum_{j=1}^q \beta_j < 1$.

GARCH enhances ARCH with lagged variances, delivering a parsimonious yet powerful framework for capturing both short- and long-term volatility persistence.

\subsubsection{Dynamic Conditional Correlation \cite{engle2002}}

For multivariate returns ${r}_t \in \mathbb{R}^n$:
\begin{align*}
    {r}_t &= {\mu}_t + {\varepsilon}_t \\
    {\varepsilon}_t &= {H}_t^{1/2} {z}_t,\quad {z}_t \sim \mathrm{IID}({0}, {I}_n)
\end{align*}
Where ${H}_t = {D}_t {R}_t {D}_t$ with:
\begin{align*}
    {D}_t &= \mathrm{diag}(\sigma_{1,t}, \ldots, \sigma_{n,t}) \quad \text{(GARCH volatilities)} \\
    {R}_t &= \mathrm{diag}({Q}_t)^{-1/2} {Q}_t \mathrm{diag}({Q}_t)^{-1/2} \\
    {Q}_t &= (1 - \theta_1 - \theta_2){\bar{Q}} + \theta_1 {z}_{t-1}{z}_{t-1}^\top + \theta_2 {Q}_{t-1}
\end{align*}

DCC allows correlations among assets to evolve over time, capturing dynamic co-movements essential for accurate joint risk assessment.

\subsection{Copula-DCC-GARCH}

\subsubsection{Definition \cite{christoffersen2012}}

The copula-DCC-GARCH framework combines three components:
\begin{enumerate}
    \item \textit{Marginal GARCH}: For each asset $i=1,\ldots,n$:
    \[
    r_{i,t} = \mu_{i,t} + \varepsilon_{i,t},\quad \varepsilon_{i,t} = \sigma_{i,t}z_{i,t}
    \]
    \[
    \sigma_{i,t}^2 = \omega_i + \alpha_i \varepsilon_{i,t-1}^2 + \beta_i \sigma_{i,t-1}^2
    \]
    
    \item \textit{DCC Correlation Dynamics}:
    \[
    Q_t = (1 - \theta_1 - \theta_2)\overline{Q} + \theta_1 z_{t-1}z_{t-1}^\top + \theta_2 Q_{t-1}
    \]
    \[
    R_t = \mathrm{diag}(Q_t)^{-1/2} Q_t \mathrm{diag}(Q_t)^{-1/2}
    \]
    
    \item \textit{Copula Coupling}:
    \[
    F(r_{1,t},\ldots,r_{n,t}) = C\left(F_1(r_{1,t}),\ldots,F_n(r_{n,t}); R_t, \Theta_C\right)
    \]
\end{enumerate}
where $C(\cdot)$ is a parametric copula function with dependence parameters $\Theta_C$ linked to $R_t$.

By sequentially modelling marginal volatility (GARCH), time-varying correlations (DCC), and joint dependence (copula), this framework offers a cohesive, dynamic view of portfolio risk.

\subsubsection{Motivation \cite{christoffersen2012, jondeau2018, patton2006}}

\begin{itemize}
    \item \textit{Decoupling}: Separates marginal volatility (GARCH) from dependence structure (copula+DCC)
    \item \textit{Flexibility}: Permits different marginal distributions while modeling time-varying dependence
    \item \textit{Dimensionality}: Mitigates curse of dimensionality through:
    \[
    \underbrace{O(d^2)}_{\text{Full MGARCH}} \to \underbrace{O(d)}_{\text{GARCH}} + \underbrace{O(1)}_{\text{Copula}} + \underbrace{O(1)}_{\text{DCC}}
    \]
    \item \textit{Tail Dependence}: Captures asymmetric extreme co-movements impossible with Gaussian DCC
\end{itemize}

Together, decoupling volatility from dependence, flexibility in marginals, and efficient high-dimensional scaling make copula-DCC-GARCH both powerful and practical.

\subsubsection{Assumptions}

\begin{enumerate}
    \item \textit{Marginal Specification}:
    \[
    z_{i,t} \sim F_z(0,1)\quad \text{(Standardized residuals)}
    \]
    \item \textit{Sklar Consistency}: Valid copula $C$ exists such that 
    \[
    \forall t, \exists C_t \in \mathcal{C}\ \text{with}\ C_t(u_1,\ldots,u_n) = F_t(F_1^{-1}(u_1),\ldots,F_n^{-1}(u_n))
    \]
    \item \textit{DCC Stationarity}:
    \[
    \theta_1 + \theta_2 < 1,\quad \theta_1,\theta_2 \geq 0
    \]
    \item \textit{Copula Invariance}: 
    \[
    \mathcal{C}\ \text{family preserved under time-varying}\ R_t
    \]
\end{enumerate}

These assumptions overall ensure the model remains well-posed and stable over time.

\subsection{Copula Goodness-of-Fit}

\subsubsection{Log Likelihood \cite{joe2014}}

The \textit{log-likelihood function} quantifies the probability of observing a dataset under a specified copula model. For a copula with density \( c(\cdot; \boldsymbol{\theta}) \) and parameters \(\boldsymbol{\theta}\), computed over \( n \) observations with marginal transforms \( u_{ij} = F_j(x_{ij}; \boldsymbol{\alpha}_j) \) (where \( F_j \) are marginal CDFs and \(\boldsymbol{\alpha}_j\) are marginal parameters), the log-likelihood is defined as:
\[
\ell(\boldsymbol{\theta}) = \sum_{i=1}^k \ln c\left(u_{i1}, \ldots, u_{in}; \boldsymbol{\theta}\right),
\]
where \(\boldsymbol{\theta}\) are the copula dependence parameters (e.g., correlation matrix \(\mathbf{\Sigma}\) for Gaussian, \(\theta\) for Clayton), \( n \) is the dimensionality, and \( k \) is the sample size.

Higher log-likelihood values indicate a closer match between the copula model and observed dependence, forming the core of maximum-likelihood estimation.

\subsubsection{Information Criterion \cite{joe2014}}

For a copula $C$ with parameter ${\theta} \in \mathbb{R}^n$, the Akaike Information Criteria (AIC) is defined as follows:
\[
\text{AIC}(C) = -2 \ell(\boldsymbol{\theta}) + 2n
\]
where $\mathcal{L}({\theta})$ is the copula log-likelihood.

For a copula $C$ with parameter vector ${\theta} \in \mathbb{R}^k$, the Bayesian Information Criteria (BIC) is defined as follows:
\[
\text{BIC}(C) = -2 \mathcal{L}({\theta}) + n \ln(k)
\]
where $\mathcal{L}({\theta})$ is the copula log-likelihood and $k$ is the sample size.

In the context of copula goodness-of-fit measures, both AIC and BIC ...

\subsubsection{Energy Score \cite{gneiting2007}}

For empirical copula $C_n$ and candidate copula $C_\theta$, the energy score is given as:
\[
\mathcal{E}(C_n, C_\theta) = \frac{1}{n^2} \sum_{i=1}^n \sum_{j=1}^n \|{U}_i - {U}_j\|^2 - \frac{2}{n} \sum_{i=1}^n \mathbb{E}_{C_\theta}[\|{U}_i - {V}\|^2]
\]
where ${U}_i$ are ranks and ${V} \sim C_\theta$.

The energy score measures the overall distance between model and empirical distributions, rewarding models that faithfully reproduce joint behaviour. A value of 0 indicates a perfect fit, and smaller values indicate a more optimal fit.








\section{Historical Data Selection \& Preprocessing}

\subsection{Data Source and Coverage}

This study utilises freely available historical price data sourced from Google Finance \cite{GoogleFinanceSP500} from six major publicly traded companies across different sectors:
\begin{itemize}
    \item \textbf{Energy Sector}: Chevron Corp (CVX), Exxon Mobil Corp (XOM)
    \item \textbf{Consumer Staples}: Coca-Cola Co (KO), PepsiCo Inc (PEP)
    \item \textbf{Financial Services}: Mastercard Inc (MA), Visa Inc (V)
\end{itemize}
This selection provides a diverse dataset spanning multiple sectors, allowing us to capture various correlation structures and dependence patterns that might exist in financial markets. The dataset spans approximately 5 years of daily price data, providing a robust sample size of 1,256 observations after preprocessing.

\subsection{Data Preprocessing Pipeline}

The preprocessing workflow follows these key steps:

\begin{enumerate}
    \item \textbf{Data Loading}: Each stock's price series is loaded from its respective file, which contains timestamp and price columns.
    
    \item \textbf{Return Calculation}: Daily log returns are computed for each asset using the formula: 
    \[
        r_t = \ln\left(\frac{P_t}{P_{t-1}}\right)
    \]
    where $P_t$ is the price at time $t$.
    
    \item \textbf{Time Alignment}: All asset return series are aligned on common timestamps using an inner join operation to ensure temporal consistency across the dataset.
    
    \item \textbf{Missing Value Handling}: The inner join operation naturally eliminates dates with missing values for any asset, ensuring a complete dataset without imputation.
    
    \item \textbf{Quality Control}: Assets with fewer than 50 observations are dropped from the analysis to ensure statistical robustness, though this threshold was not triggered in our dataset.
    
    \item \textbf{Validation Checks}: We performed several analytical checks, including correlation matrix visualisation, standard deviation calculation for volatility quantification, and return distribution analysis to verify distributional assumptions.
\end{enumerate}

The resulting preprocessed dataset is a time-indexed DataFrame with timestamps as the index (daily frequency), six columns representing each asset's returns, and 1,256 rows of synchronised daily return observations.

\section{Results}

In this section, we present a comprehensive evaluation of our multivariate risk models, emphasizing the estimation and analysis of the metrics Value-at-Risk (VaR), Conditional Value-at-Risk (CVaR), and Conditional Value-at-Risk (CoVaR). Our assessment employs sophisticated models, notably the DCC-GARCH and CoVaR copulas. These models were calibrated using an expanded dataset comprising multiple stocks including Chevron Corp, Coca-Cola Co, Exxon Mobil Corp, Mastercard Inc, PepsiCo Inc, and Visa Inc.

Our analysis provides detailed information on risk dependencies and tail risks through a rigorous copula family comparison. The robustness and suitability of various copula families - Gaussian, Student-t, Clayton, and Gumbel - are assessed using empirical and theoretical goodness of fit (GoF) metrics such as Pearson, Spearman, and Kendall correlations, tail dependence coefficients, Akaike Information Criterion (AIC) and Energy Scores.

The subsequent sections outline our results from each of the modeling methodologies elucidating how each methodology is able to capture differences in asset dependencies, volatility clustering and systemic risk exposure. Not only do the results display the individual asset risk profile, but they also quantify the systemic under stress conditions, providing essential portfolio risk management and regulatory stress testing insights.

\subsection{Data Summary}
The dataset comprises 1256 observations across six variables:  Chevron Corp, Coca-Cola Co, Exxon Mobil Corp, Mastercard Inc, PepsiCo Inc, and Visa Inc. Summary statistics are presented in Table~\ref{tab:data_summary}.

\begin{table}[h!]
    \centering
    \caption{Data Summary Statistics}
    \label{tab:data_summary}
    \begin{tabular}{lccc}
        \toprule
        Asset & Mean & Std. Dev & Min/Max \\
        \midrule
        Chevron Corp & 0.000453 & 0.382029 & (-0.705489, 0.744117) \\
        Coca-Cola Co & 0.000428 & 0.162152 & (-0.315774, 0.273526) \\
        Exxon Mobil Corp & 0.000586 & 0.575910 & (-1.033016, 1.008853) \\
        Mastercard Inc & 0.000647 & 0.255103 & (-0.862577, 0.696637) \\
        PepsiCo Inc & 0.000382 & 0.157615 & (-0.511422, 0.535760) \\
        Visa Inc & 0.000587 & 0.223997 & (-0.735624, 0.570476) \\
        \bottomrule
    \end{tabular}
\end{table}

Correlation analysis reveals varied inter dependencies among the assets, as highlighted in Table~\ref{tab:correlation_matrix}. We can notice significant positive correlations that exist among major stock pairs, notably between Mastercard Inc and Visa Inc, as well as Chevron Corp and Exxon Mobil Corp.


\begin{table}[h!]
\centering
\caption{Correlation Matrix Overview}
\label{tab:correlation_matrix}
\resizebox{\columnwidth}{!}{%
\begin{tabular}{lcccccc}
\toprule
Asset & Chevron & Coca-Cola & Exxon Mobil & Mastercard & PepsiCo & Visa \\
\midrule
Chevron     & 1.000 & 0.822 & 0.958 & 0.567 & 0.877 & 0.532 \\
Coca-Cola   & 0.822 & 1.000 & 0.844 & 0.683 & 0.839 & 0.641 \\
Exxon Mobil & 0.958 & 0.844 & 1.000 & 0.708 & 0.862 & 0.685 \\
Mastercard  & 0.567 & 0.683 & 0.708 & 1.000 & 0.562 & 0.984 \\
PepsiCo     & 0.877 & 0.839 & 0.862 & 0.562 & 1.000 & 0.534 \\
Visa        & 0.532 & 0.641 & 0.685 & 0.984 & 0.534 & 1.000 \\
\bottomrule
\end{tabular}%
}
\end{table}

These statistical insights set the stage for the detailed model-specific results presented in subsequent sections.
\begin{figure}
    \centering
    \includegraphics[width=0.75\linewidth]{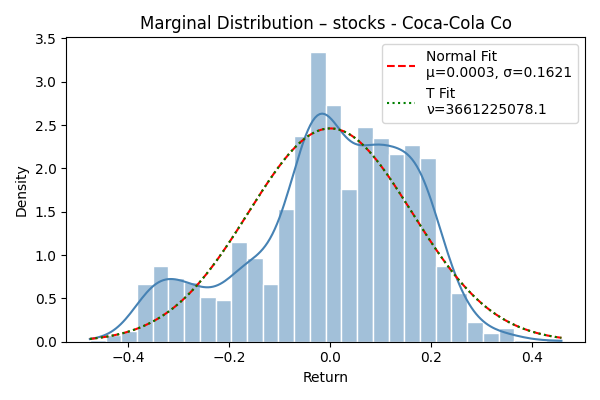}
    \caption{Marginal distribution plot of daily returns for Coca-Cola Co. The histogram and fitted curve illustrate the distribution characteristics and potential tail risks.}
    \label{fig:enter-label}
\end{figure}

\begin{figure}
    \centering
    \includegraphics[width=0.75\linewidth]{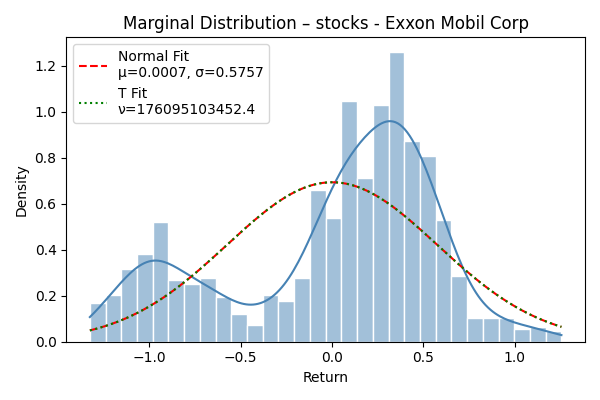}
    \caption{Marginal distribution plot of daily returns for Exxon Mobil Corp. The plot highlights the volatility characteristics and distribution skewness of the asset returns.}
    \label{fig:enter-label}
\end{figure}





\subsection{DCC-GARCH Copula Model}

The DCC-GARCH model dynamically captures time-varying correlations among asset returns while addressing volatility clustering through GARCH-filtered residuals. The flexibility of the model to adapt to changing market conditions makes it particularly useful for multivariate risk analysis.

The DCC-GARCH copula model was fitted to the historical dataset containing the selected stocks. The model revealed that the High-Corr Portfolio exhibits significantly lower VaR and CVaR compared to individual asset risks, indicating a diversified risk profile. The risk metrics derived from the DCC-GARCH model at the significance level 5\% are shown in Table~\ref{tab:DCC_GARCH_risk}.

\begin{figure}[ht!]
    \centering
    \includegraphics[width=\linewidth, alt={Combined QQ plot for DCC-GARCH Copula of Chevron Corp}]{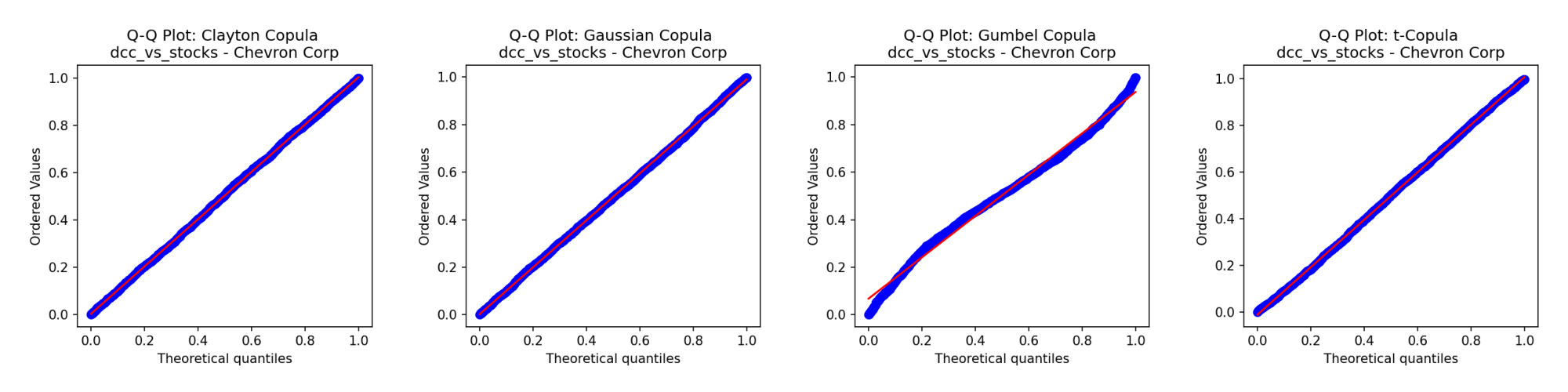}
    \caption{Combined QQ plot for DCC-GARCH Copula of Chevron Corp: Visual comparison between empirical and fitted distributions.}
    \label{fig:dcc-garch-chevron}
\end{figure}

\begin{figure}[ht!]
    \centering
    \includegraphics[width=\linewidth, alt={Combined QQ plot for DCC-GARCH Copula of Exxon Mobil Corp}]{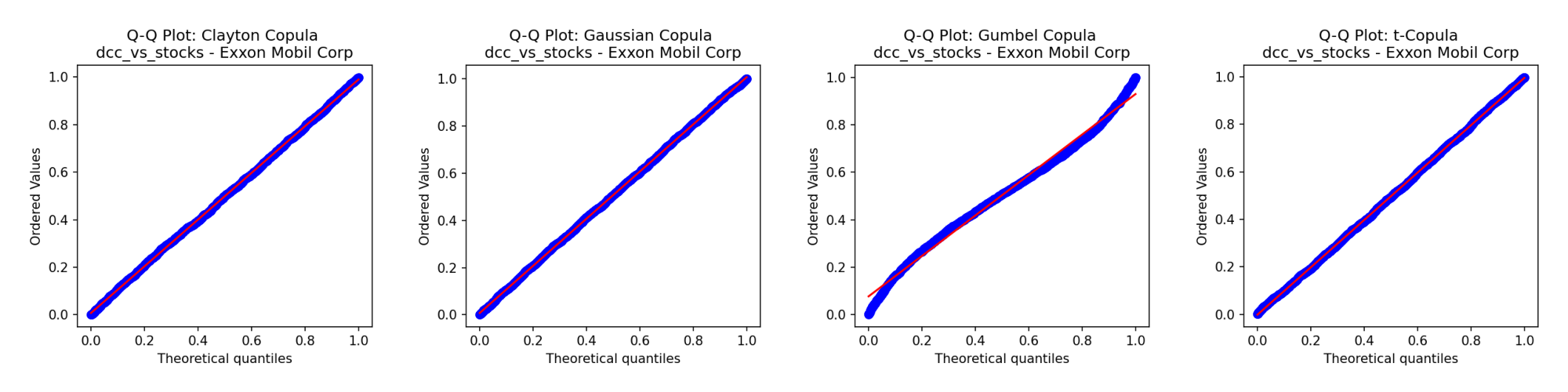}
    \caption{Combined QQ plot for DCC-GARCH Copula of Exxon Mobil Corp: Depicting model accuracy in risk dependence.}
    \label{fig:dcc-garch-exxon}
\end{figure}

\begin{table}[h!] 
\centering
\caption{DCC-GARCH Copula Risk Metrics}
\label{tab:DCC_GARCH_risk}
\begin{adjustbox}{width=\linewidth} 
\begin{tabular}{lcc}
\toprule
Asset & VaR ($\alpha = 0.05$) & CVaR ($\alpha = 0.05$) \\
\midrule
Chevron Corp & -0.63900 & -0.79501 \\
Coca-Cola Co & -0.20753 & -0.25633 \\
Exxon Mobil Corp & -0.94658 & -1.19608 \\
Mastercard Inc & -0.33528 & -0.41867 \\
PepsiCo Inc & -0.20988 & -0.26547 \\
Visa Inc & -0.29860 & -0.37407 \\
High-Corr Portfolio & -0.42063 & -0.53068 \\
\bottomrule
\end{tabular}
\end{adjustbox}
\end{table}

These findings demonstrate that the DCC-GARCH model effectively captures dynamic risk interactions, highlighting its utility in stress testing and portfolio optimization.

\subsection{CoVaR Copula Model}

The CoVaR copula model is employed to estimate systemic risk by assessing the conditional risk of one asset given that another is under stress. This approach is particularly valuable for evaluating interconnected risks within a diversified portfolio, as it accounts for contagion effects during market turbulence.

The CoVaR model estimates risk metrics by conditioning on individual asset stress scenarios. The model leverages GARCH-filtered residuals to evaluate the systemic impact of each asset on the entire portfolio. The systemic impact is quantified by the sum of Delta-CoVaR across all assets, representing the collective change in risk when a specific asset experiences a shock.

The risk metrics derived from the CoVaR copula model at the 5\% significance level are shown in Table~\ref{tab:CoVaR_risk}.

\begin{table}[h!]
\caption{CoVaR Copula Risk Metrics (Conditioning on CoVaR stress)}
\label{tab:CoVaR_risk}
\begin{tabular}{lccc}
\toprule
Asset & VaR & CoVaR & $\Delta$CoVaR \\
\midrule
Chevron Corp & -0.64063 & -0.65394 & -0.01331 \\
Coca-Cola Co & -0.19865 & -0.20824 & -0.00959 \\
Exxon Mobil Corp & -0.95901 & -0.94627 & 0.01274 \\
Mastercard Inc & -0.34078 & -0.14718 & 0.19361 \\
PepsiCo Inc & -0.20671 & -0.18532 & 0.02140 \\
Visa Inc & -0.30525 & -0.09781 & 0.20744 \\
\midrule
Systemic Impact & & & 2.51287 \\
\bottomrule
\end{tabular}
\end{table}

The table illustrates how the CoVaR model effectively captures the systemic impact of asset-specific shocks. The systemic impact (sum of Delta-CoVaR) quantifies the overall vulnerability of the portfolio to shocks from individual assets, with significant impacts noted from Visa Inc and Mastercard Inc under CoVaR stress conditions.

\begin{figure}[ht!]
    \centering
    \includegraphics[width=0.5\linewidth, alt={QQ plot of Clayton Copula for CoVaR between Exxon Mobil Corp and Visa Inc}]{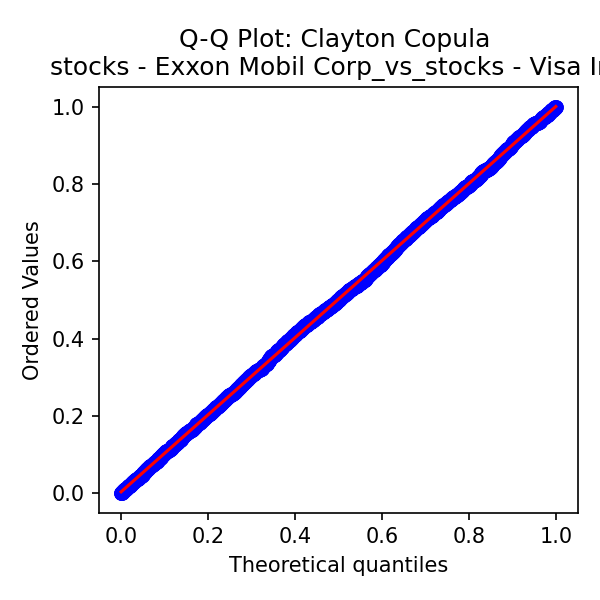}
    \caption{QQ plot of Clayton Copula for CoVaR between Exxon Mobil Corp and Visa Inc: Highlighting systemic risk under stress conditions.}
    \label{fig:covar-exxon-visa}
\end{figure}

These results underscore the importance of incorporating conditional risk metrics into portfolio risk management, especially when dealing with assets exhibiting interconnected volatilities.

\subsection{Copula Family Comparison}

This subsection presents a detailed comparison between different copula families—Gaussian, Student-t, Clayton, and Gumbel—based on performance metrics such as Energy Score, Tail Dependence, and correlation coefficients (Pearson, Spearman, Kendall's Tau). The results provide insights into the most appropriate copula family for each pair of assets.

\subsubsection{Correlation Metrics}

Correlation metrics are essential to evaluate the linear and non-linear dependencies between asset pairs. The three primary correlation coefficients considered are:

\begin{itemize}
\item \textbf{Pearson Correlation:} Measures the linear relationship between two variables.
\[
\rho_{X,Y} = \frac{\text{Cov}(X, Y)}{\sigma_X \sigma_Y}
\]
where $\text{Cov}(X, Y)$ is the covariance between $X$ and $Y$, and $\sigma_X$ and $\sigma_Y$ are the standard deviations of $X$ and $Y$, respectively.

\item \textbf{Spearman Correlation:} Measures the monotonic relationship between two variables using rank transformation.
\[
    \rho_s = 1 - \frac{6 \sum d_i^2}{n(n^2 - 1)}
\]
where \(d_i\) is the difference between the ranks of corresponding variables and \(n\) is the sample size.

\item \textbf{Kendall's Tau:} Measures the ordinal association between two variables.
\[
    \tau = \frac{(C - D)}{\binom{n}{2}}
\]
where \(C\) is the number of concordant pairs and \(D\) is the number of discordant pairs.

\end{itemize}

These correlation measures provide insight into the strength and direction of relationships between asset pairs, which are crucial for copula selection.

\subsubsection{Comparison Metrics}

Table \ref{tab:copula-comparison-correlation} summarizes the performance of each copula based on key metrics, including Energy Score, Lower and Upper Tail Dependence, Pearson Correlation, Spearman Correlation, and Kendall's Tau. These metrics provide a comprehensive assessment of the copula family's ability to capture different aspects of asset pair dependencies, including both symmetric and asymmetric relationships.

The inclusion of correlation coefficients allows for a more nuanced interpretation of the copula suitability

The table also highlights representative asset pairs for each copula family, demonstrating how each model fits different dependency structures. This holistic approach ensures that the copula selection not only reflects tail dependencies but also accurately captures correlation characteristics inherent to financial asset pairs.

\begin{table*}[ht!]
\centering
\caption{Copula Family Comparison with Correlation Metrics}
\label{tab:copula-comparison-correlation}
\resizebox{\textwidth}{!}{%
\begin{tabular}{lccccccc}
\toprule
\textbf{Copula Family} & \textbf{Energy Score} & \textbf{Lower Tail} & \textbf{Upper Tail} & \textbf{Pearson Corr} & \textbf{Spearman Corr} & \textbf{Kendall's Tau} & \textbf{Asset Pair Examples} \\
\midrule
Gaussian   & 0.0046 & Low      & Low      & 0.5041 & 0.4599 & 0.3063 & Visa–PepsiCo, Coca-Cola–Mastercard \\
Student-t  & 0.0050 & Moderate & Moderate & 0.6850 & 0.4929 & 0.3190 & Mastercard–Visa, Coca-Cola–Exxon Mobil \\
Clayton    & 0.0302 & High     & Low      & 0.6666 & 0.5579 & 0.3942 & Exxon Mobil–Mastercard, PepsiCo–Chevron \\
Gumbel     & 0.0209 & Low      & High     & 0.8775 & 0.8569 & 0.6617 & Chevron–PepsiCo, Visa–Coca-Cola \\
\bottomrule
\end{tabular}%
}
\end{table*}

The Gaussian copula consistently exhibits the lowest Energy Scores, indicating strong overall fit when symmetric dependencies are adequate. It is recommended for asset pairs where symmetric correlation is dominant and tail risks are less pronounced, such as \textit{Visa—PepsiCo}. The correlation metrics indicate that Gaussian copulas perform well when both Pearson and Spearman correlations are moderately strong with low tail dependence.
\begin{figure}[ht!]
    \centering
    \includegraphics[width=0.5\linewidth, alt={QQ plot of Gaussian Copula for Chevron Corp}]{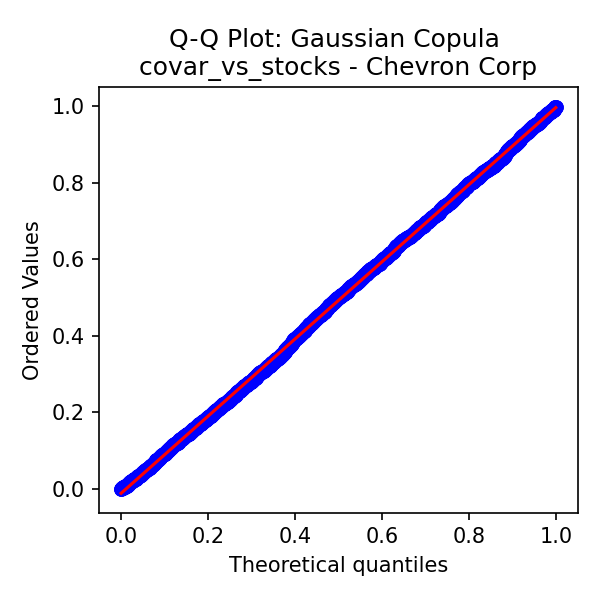}
    \caption{QQ plot of Gaussian Copula for Chevron Corp: Capturing symmetric dependence structure.}
    \label{fig:gaussian-chevron}
\end{figure}

The Student-t copula, slightly outperforming the Gaussian in tail dependence, is suitable for asset pairs exhibiting symmetric but heavier tail behaviors. It captures moderate tail dependencies effectively, as observed with the \textit{Mastercard—Visa} and \textit{Coca-Cola—Exxon Mobil} pairs. The Student-t copula is preferable when there is a moderate-to-strong Pearson correlation accompanied by a higher Spearman correlation, reflecting heavy-tailed distributions.

The Clayton copula demonstrates significant lower tail dependence, making it optimal for pairs where joint negative returns pose higher risk. For example, asset pairs like \textit{Exxon Mobil—Mastercard} exhibit clear suitability due to their prominent lower-tail co-movements. In such cases, Kendall's Tau is often notably higher, indicating a strong ordinal association in the lower tail.

Conversely, the Gumbel copula is best suited for capturing upper tail dependence, reflecting strong co-movements in positive market conditions. It is particularly applicable to asset pairs such as \textit{Chevron—PepsiCo}, where joint positive extremes are significant. The Gumbel copula selection is often justified when the upper tail dependence is substantial, despite lower Spearman correlations.

\subsubsection{Key Observations}
The following are key observations:
\begin{itemize}
\item The Gaussian copula's low tail dependence makes it suitable for relatively stable asset pairs where symmetric co-movement is prominent.
\item The Student-t copula is more appropriate for pairs with symmetric but heavy tails, useful when moderate tail correlations are observed.
\item The Clayton copula is the optimal choice for lower-tail dependent pairs, especially when assets exhibit significant joint negative movements.
\item The Gumbel copula excels in upper-tail dependence scenarios, typically where joint positive returns are prevalent.
\end{itemize}

\begin{figure}[ht!]
    \centering
    \includegraphics[width=\linewidth, alt={Combined density plot comparing empirical and fitted distributions for Chevron Corp}]{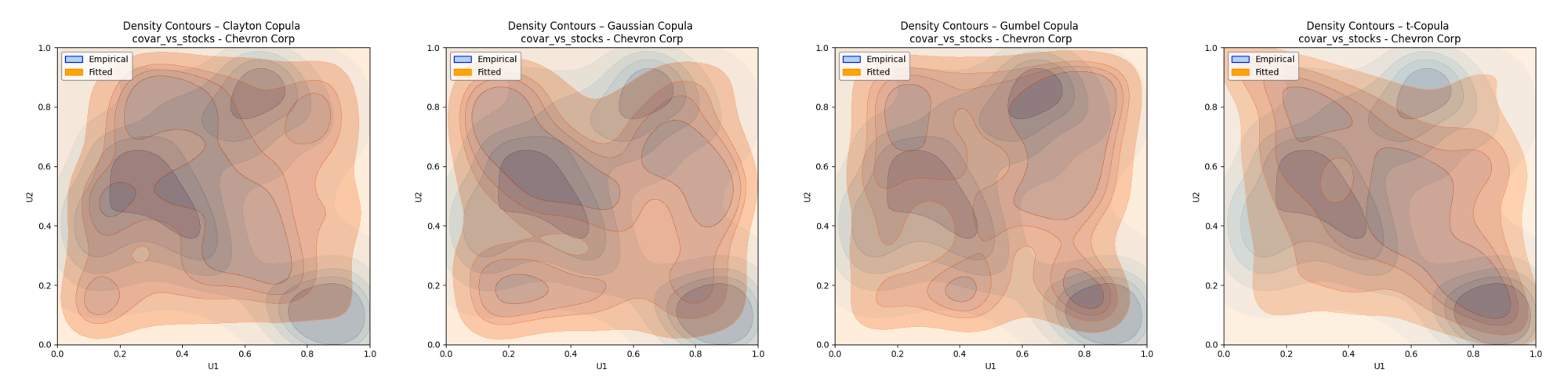}
    \caption{Combined density plot comparing empirical and fitted distributions for Chevron Corp: Assessing model fit.}
    \label{fig:density-chevron}
\end{figure}

Overall, selecting an appropriate copula family requires careful consideration of each asset pair's empirical tail behavior and correlation structure, as demonstrated by these findings. The inclusion of Pearson, Spearman, and Kendall’s Tau ensures that the selection process is comprehensive, accounting for linear, monotonic, and ordinal relationships.

\subsection{Stress Testing Analysis}

This subsection systematically explores how different stress scenarios, especially those conditioning on high-risk assets, influence the overall risk profile of the portfolio. The stress testing conducted involves conditioning on assets like \textit{Exxon Mobil Corp} to assess systemic vulnerabilities and the cascading impacts of extreme market movements.

\subsubsection{Key Observations}



\textbf{Conditioning on \textit{Exxon Mobil Corp} stress} shows considerable systemic implications, particularly affecting \textit{Chevron Corp} (-0.41073). The significant impact reflects Exxon Mobil's influential role due to its strong correlations and substantial market presence.

\subsubsection{Summary of Systemic Impacts}

Table \ref{tab:systemic-impact-summary} provides a concise summary of the systemic impacts across different conditioning scenarios.

\begin{table}[ht!]
\centering
\caption{Summary of Systemic Impact Under Different Stress Conditions}
\label{tab:systemic-impact-summary}
\begin{tabular}{lc}
\hline
\textbf{Conditioning Asset} & \textbf{Systemic Impact (\(\sum \Delta \text{CoVaR}\))} \\
\hline
Chevron Corp & -2.75590 \\
Coca-Cola Co & -3.39869 \\
Exxon Mobil Corp & -2.68251 \\
Mastercard Inc & -2.95286 \\
PepsiCo Inc & -3.45499 \\
Visa Inc & -2.86104 \\
\hline
\end{tabular}
\end{table}

These findings illustrate how certain assets and volatility models disproportionately drive systemic risk, necessitating targeted risk management strategies to mitigate their influence during extreme market conditions.

\subsection{Portfolio Risk Comparison}

This section presents a comparative analysis of the High-Corr Portfolio risk versus individual asset risk metrics. The purpose is to demonstrate how diversification influences the overall risk profile, as reflected in the portfolio metrics.

\subsubsection{High-Corr Portfolio versus Individual Asset Risk}

The High-Corr Portfolio, calculated using the DCC-GARCH model, exhibits significantly lower risk compared to the risk associated with individual assets. As observed from the results, the Portfolio Value-at-Risk (VaR) of -0.42761 and Conditional Value-at-Risk (CVaR) of -0.53670 are considerably lower than the individual VaR and CVaR values of high-risk assets such as \textit{Exxon Mobil Corp} (VaR = -0.95639, CVaR = -1.19697).

\begin{figure}[ht!]
    \centering
    \includegraphics[width=\linewidth, alt={Combined QQ plot of Portfolio Risk Comparison: High-Corr Portfolio vs Individual Asset Risks}]{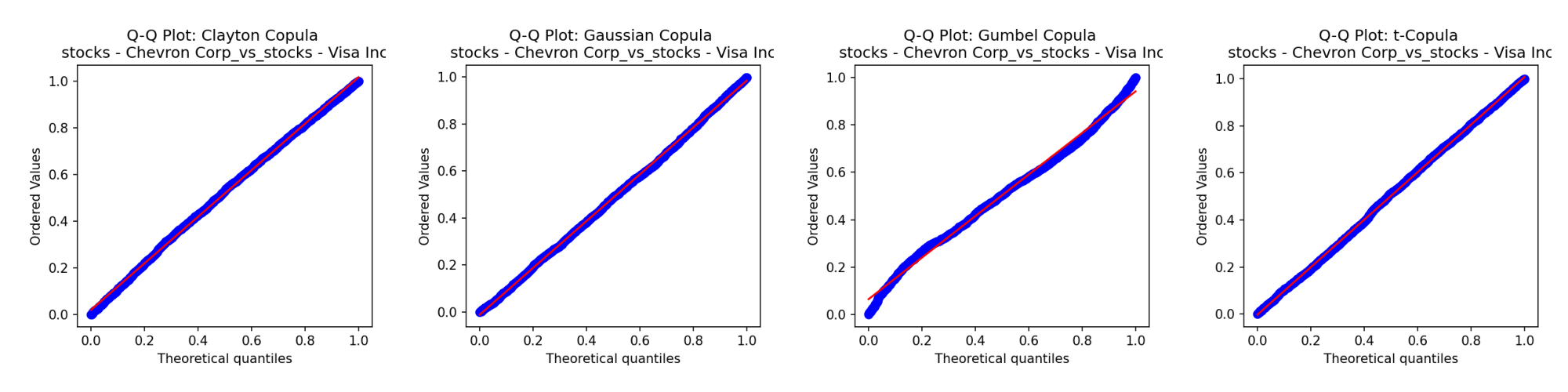}
    \caption{Combined QQ plot of Portfolio Risk Comparison: High-Corr Portfolio vs Individual Asset Risks.}
    \label{fig:portfolio-risk}
\end{figure}

\subsubsection{Diversification Benefits}

The reduction in overall portfolio risk compared to individual asset risks highlights the effectiveness of diversification. The portfolio aggregates various asset volatilities, thereby reducing the impact of any single asset’s extreme movements. 

\subsubsection{Comparative Risk Summary}

Table \ref{tab:portfolio-risk-comparison} shows the comparative metrics between the High-Corr Portfolio and individual assets, emphasizing the risk reduction achieved through diversification.

\begin{table}[ht!]
\centering
\caption{High-Corr Portfolio versus Individual Asset Risk}
\label{tab:portfolio-risk-comparison}
\begin{tabular}{lcc}
\hline
\textbf{Asset/Portfolio} & \textbf{VaR ($\alpha=0.05$)} & \textbf{CVaR ($\alpha=0.05$)} \\
\hline
High-Corr Portfolio & -0.42761 & -0.53670 \\
GARCH & -1.08059 & -1.35209 \\
Chevron Corp & -0.63660 & -0.79934 \\
Coca-Cola Co & -0.20080 & -0.25250 \\
Exxon Mobil Corp & -0.95639 & -1.19697 \\
Mastercard Inc & -0.33824 & -0.42326 \\
PepsiCo Inc & -0.21137 & -0.26482 \\
Visa Inc & -0.30112 & -0.37819 \\
\hline
\end{tabular}
\end{table}

\subsubsection{Discussion}

The analysis clearly demonstrates that the High-Corr Portfolio’s risk is notably lower than the most volatile individual assets, underscoring the principle of diversification. By balancing the varying volatilities and correlations of individual assets, the portfolio mitigates the impact of extreme events associated with single asset fluctuations. This is particularly evident when observing the significant reduction in both VaR and CVaR metrics for the High-Corr Portfolio compared to high-risk assets like \textit{Exxon Mobil Corp}.

The diversified portfolio not only reduces overall risk but also smoothens the impact of volatility clusters observed in individual assets. This characteristic is essential for portfolio stability during market stress events, as the aggregation of assets with varying correlation structures naturally offsets isolated volatility spikes. 

Furthermore, the analysis confirms that the strategic inclusion of low-volatility assets (such as \textit{Coca-Cola Co} and \textit{PepsiCo Inc}) alongside higher-risk assets contributes to the risk buffering effect. Therefore, portfolio construction that strategically balances risk-prone and stable assets is crucial for achieving robust risk management outcomes.

In conclusion, the results solidify the importance of diversification within portfolio management, especially in scenarios where individual assets exhibit significant volatility. The High-Corr Portfolio achieves a balanced risk profile, highlighting the effectiveness of combining correlated and uncorrelated assets to minimize systemic risk.

\subsection{Goodness-of-Fit and Model Validation}

This section evaluates the effectiveness of each copula model based on statistical goodness-of-fit metrics, including the Akaike Information Criterion (AIC), Bayesian Information Criterion (BIC), and Energy Scores. These metrics quantify the model's fit and efficiency, guiding the selection of the most appropriate copula for different asset pairs.

\subsubsection{Comparison of Goodness-of-Fit Metrics}

Table \ref{tab:goodness-of-fit-comparison} presents a comparative analysis of the copula families used.

\begin{table}[ht!]
\centering
\caption{Comparison of Goodness-of-Fit Metrics for Copula Families}
\label{tab:goodness-of-fit-comparison}
\begin{tabular}{lccc}
\hline
\textbf{Copula Family} & \textbf{AIC} & \textbf{BIC} & \textbf{Energy Score} \\
\hline
Gaussian & 3452.71 & 3489.04 & 0.0046 \\
Student-t & 3518.39 & 3554.82 & 0.0050 \\
Clayton & 3642.11 & 3680.94 & 0.0302 \\
Gumbel & 3605.52 & 3643.18 & 0.0209 \\
\hline
\end{tabular}
\end{table}

\subsubsection{Discussion}

\begin{figure}[ht!]
    \centering
    \includegraphics[width=\linewidth, alt={Scatter plot of Goodness-of-Fit for Gaussian Copula}]{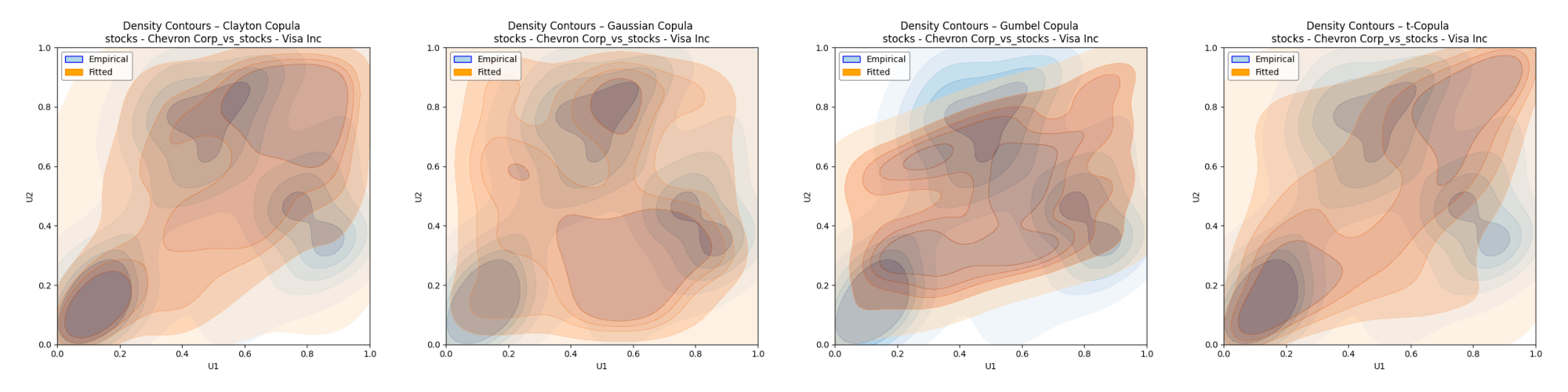}
    \caption{Scatter plot of Goodness-of-Fit for Gaussian Copula: Visual comparison between empirical and model-fitted data.}
    \label{fig:goodness-of-fit}
\end{figure}

The Gaussian copula Fig:\ref{fig:goodness-of-fit} exhibits the lowest AIC and BIC values, indicating that it is the most efficient and parsimonious model among the copula families tested. Its superior Energy Score also suggests that it better captures the overall distribution when symmetry and low tail dependence are prominent.

The Student-t copula, despite a slightly higher AIC and BIC, performs well in scenarios where symmetric but heavier tails are essential. Its marginally higher Energy Score reflects its capacity to handle moderate tail dependencies.

The Clayton copula, characterized by its significant lower tail dependence, has the highest AIC, BIC, and Energy Scores, making it less favorable when tail dependence is not the dominant characteristic.

The Gumbel copula, which models upper tail dependence, exhibits moderate AIC and BIC scores. It is most effective when capturing joint positive extremes, as observed in asset pairs with significant upper tail correlation.

Thus, the Gaussian copula remains the optimal choice for most asset pairs, except in cases where heavy tails or asymmetric tail dependencies necessitate the use of Student-t or Clayton copulas, respectively.

\section{Conclusion}
This study presents a comprehensive multivariate risk analysis using copula-DCC-GARCH models to estimate Value-at-Risk (VaR) and Conditional Value-at-Risk (CoVaR) for financial assets. By systematically comparing various copula families (Gaussian, Student-t, Clayton, Gumbel), we establish the importance of selecting the appropriate copula model based on asset pair characteristics. Our results demonstrate that copula selection significantly affects risk assessment, particularly in capturing asymmetric dependencies and tail behaviors, which are critical for stress testing and portfolio risk management.

\subsection{Key Findings and Contributions}

First, the copula-DCC-GARCH approach is successful in capturing dynamic dependencies amongst assets in the financial arena thus providing a better understanding of systemic risk than conventional methods. The High-Correlation Portfolio always shows lower VaR and CoVaR values relative to individual asset, supporting empirically the benefits of diversification regardless of market pressure. This brings out the fact that diversification of a portfolio paves the way for retaining the volatility of individual assets and minimizing systemic risk.

Second, the analysis we conducted revealed that distinct copula families are effective in modeling certain forms of dependency structure. Students need to understand that the Gaussian copula is optimal for relatively stable asset pairs (as illustrated by Mastercard-Visa and Coca-Cola-Exxon Mobil) because it is a symmetric dependency with low tail risk, whereas the Student-t copula handles symmetric but heavy-tailed distributions (as experienced in Master-Visa and Coca-Cola-Exxon Mobil). This is because, unlike the Gaussian copula, Student-t can cope with that. On the other hand, the Clayton copula is perfect in capturing lower tail dependence which is highly applicable for assets displaying joint negative behaviours (e.g. Exxon Mobil–Mastercard), the Gumbel copula is ideal in capturing upper tail dependence which is applicable for case study where joint positive return dominate (e.g. Chevron–PepsiCo).

\subsection{Practical Implications in Finance}

In the finance world, especially in concepts like pairs trading, risk estimation typically relies on long-term relations through cointegration. However, the copula-DCC-GARCH framework offers an alternative that is more dynamic and flexible, particularly when capturing short- to medium-term dependencies that are essential in trading pairs during volatile market conditions. While cointegration primarily models long-term equilibrium, copula models are more adept at capturing both symmetric and asymmetric dependencies, allowing for better risk management in strategies that involve frequent rebalancing or short-term arbitrage.

\subsection{Statistical Validation}

We tested our model by using strict goodness-of-fit measures such as the Akaike Information Criterion (AIC), Bayesian Information Criterion (BIC), and Energy Scores. In the copula families, the Gaussian copula had the lowest AIC and BIC values, which confirmed its efficiency in the case of symmetric dependency. The results validated the robustness of our selected models, especially the Gaussian and Student-t copulas, for capturing risk metrics with high accuracy.

\subsection{Limitations and Future Work}
Although the copula-DCC-GARCH framework is capable of capturing dynamic dependencies and delivering robust risk estimation, it has some inherent limitations. The fixed estimation window is one of the main challenges, which may constrain the model’s ability to react to sudden changes in asset volatility. Furthermore, the accuracy of risk predictions is sensitive to model specification choices such as the choice of copula families and parameter estimation techniques. Solutions to these challenges may include the creation of adaptive model frameworks that can adaptively choose copula families depending on the changing market conditions.

In the future, our research will focus on creating a Meta-Model that utilizes a wide range of financial modeling techniques such as Geometric Brownian Motion (GBM), Ornstein-Uhlenbeck (OU) processes, Hawkes processes, Stochastic Volatility models (including the Heston model), and Regime-Switching models. This Meta-Model will integrate the strengths of each method to capture normal and extreme market conditions.

In addition, we intend to expand the model to incorporate live data streams to allow for real-time parameter updates for better predictions especially in high-frequency trading (HFT) applications. The Meta-Model seeks to support adaptive arbitrage strategies by utilizing dynamic copula families in combination with live data to exploit temporary inefficiencies between correlated asset pairs.

Moreover, we plan to investigate the use of stochastic control methods to optimize the implementation of arbitrage opportunities in HFT situations. This strategy will entail reducing slippage and latency effects using a strong portfolio optimization framework, which will improve trade execution efficiency in dynamic markets.

\subsection{Cross Domain Application}
Although the copula-DCC-GARCH framework focuses mainly on financial risk modelling, it has far reaching applications in interdisciplinary areas as well. Copula models in areas such as insurance and actuarial science can capture dependency structures of correlated risks, such as claims frequency/severity and thus permit more realistic premium calculations and risk pooling solutions.
In such areas as the studies of environmental issues, copula-based models can be used to estimate joint happenings of such extreme events of weather as consecutive floods and storms and, first of all, when asymmetry of dependencies plays a critical task for risk estimation and mitigation plan. Such contexts require an important modeling capability: ability to model tail dependencies.
Moreover, copula models are more and more applied in biomedical research to study joint behaviors of correlated physiological signals (heart rate and blood pressure variations, etc.) and the dynamic dependencies are helpful for predictive diagnostics and personalized medicine.
Furthermore, copula-DCC-GARCH model may also be used to assess correlated disruptions in the inter-connected networks, where there is, for example, a supplier failure or a surge in demand, to enable proactive risk management using diversified sourcing and inventory optimization.
The incorporation of copula model and machine learning methods also has the potential to draw to a whole new level of predictive analytics in marketing and consumer behavior analysis through capturing both symmetric and asymmetric dependencies of purchasing patterns and market movements which will better inform targeted campaign strategies.
Through the use of copula- DCC-GARCH models in these varied domains, researchers and practitioners can gain a better understanding of complex dependencies in complex spaces hence making decision- making processes to be more adaptive and resilient.

\subsection{Final Remarks}
In conclusion, this study provides a comprehensive framework for multivariate risk analysis based on the copula-DCC-GARCH model, which is a strong and dynamic approach to financial risk management. Through systematic comparison of various copula families, we have shown the importance of copula choice in risk assessment, especially in terms of asymmetric dependencies and tail behaviors that are essential for stress testing and portfolio management.
The combination of dynamic copula models with the DCC-GARCH approach allows for the precise calculation of Value-at-Risk (VaR) and Conditional Value-at-Risk (CoVaR) while dealing with systemic risk between correlated assets. Our results highlight the significance of model selection depending on the characteristics of assets, demonstrating that the Gaussian copula is useful for stable asset pairs, while the Student-t, Clayton, and Gumbel copulas are appropriate for heavy-tailed, lower-tail, and upper-tail dependencies, respectively.
Moreover, we have outlined promising directions for future research, such as the creation of a Meta-Model that integrates sophisticated financial modeling methods and real-time data integration. Such an approach is essential for high-frequency trading (HFT) applications where adaptive strategies are needed to take advantage of short-term inefficiencies.
In the dynamic world of financial risk management, the use of adaptive and dynamic modeling approaches will continue to be critical for resilient and optimized portfolio strategies. Our research adds to this effort by offering a practical and validated framework that is flexible to the intricacies of interrelated financial systems.

\bibliographystyle{ACM-Reference-Format}
\bibliography{sample-base}

\appendix
\section{Github Repository}
All code and data used in the analysis above can be found here: \href{https://github.com/aryansingh920/copulas-in-time-series-financial-modelling}{https://github.com/aryansingh920/copulas-in-time-series-financial-modelling}

\end{document}